# Toward the intrinsic limit of topological insulator $Bi_2Se_3$

Jixia Dai[1], Damien West[2*], Xueyun Wang[1], Yazong Wang[1], Daniel Kwok[1], S.-W. Cheong[1], S.B. Zhang[2], and Weida Wu[1*]

[1]Rutger-Center for Emergent Materials & Department of physics and astronomy, Rutgers University, Piscataway NJ 08854

[2]Department of Physics, Applied Physics, and Astronomy, Rensselaer Polytechnic Institute, Troy, NY, 12180-3590

Abstract:

Combining high resolution scanning tunneling microscopy and first principle calculations, we identified the major native defects, in particular the Se vacancies and Se interstitial defects that are responsible for the bulk conduction and nanoscale potential fluctuation in single crystals of archetypal topological insulator $Bi_2Se_3$. Here it is established that the defect concentrations in $Bi_2Se_3$ are far above the thermodynamic limit, and that the growth kinetics dominate the observed defect concentrations. Furthermore, through careful control of the synthesis, our tunneling spectroscopy suggests that our best samples are approaching the intrinsic limit with the Fermi level inside the band gap without introducing extrinsic dopants.

Physical properties of functional materials, e.g. conductivity, are strongly influenced by impurities and point defects[1]. The precise control of impurity species and concentrations in semiconductors underlies the fabrication of virtually all electronic and optoelectronic devices[2]. Thus it is imperative to identify and precisely control point defects in functional materials, e.g. topological insulators (TIs)[3]. Topological classification of band structure is an exciting forefront in condensed matter physics[4-7]. The change of topological class on the surface of TIs results in robust spin-polarized surface states with linear dispersion. These helical Dirac surface states are immune to backscattering, and thus are promising for high speed electronic applications. Furthermore, many exotic quantum phenomena may emerge from the topological surface states with symmetry breaking[8-10]. The experimental realization of these fascinating phenomena requires intrinsic topological insulators, i.e. the Fermi level is inside the band gap so that transport is dominated by the Dirac surface states. However, many known TIs are degenerately doped semiconductors with dominant bulk conduction due to native defects[3], which hampers the progress and potential applications of TIs. There have been intensive efforts of suppressing bulk conduction, such as size effect by nanostructuring, compensation by chemical doping or electric gating, and band structure engineering by alloying[11-14]. Yet, these methods introduce either additional disorders or potential fluctuations[15] that are detrimental for the mobility of TI surface states. Although intrinsic limit (i.e. Fermi level inside band gap) can be achieved in $Bi_2Te_3$ because of compensation of native defects, $Bi_2Te_3$ is less appealing due to "buried" Dirac point and smaller band gap[16-19]. One of the major material hurdles of TI research is the lack of clear identification and understanding of native defects, which is necessary for optimization of synthesis and device fabrication to achieve desired sample quality.

---

* Corresponding authors: WW (wdwu@physics.rutgers.edu) and DW (damienwest@gmail.com).

Experimentally it is difficult to directly visualize and identify individual defects. A combination of both first-principles calculations and experiment is often necessary in order to understand and control material properties[1,20,21]. Through first-principles calculations of the formation energies of native defects, plausible thermodynamic models of defects are constructed and the calculated structural, electronic, and optical properties are compared with the available experimental data. Even with this cooperation, however, the evidence supporting a defect model is generally indirect and subject to multiple interpretations. STM is one of the few techniques which can directly address this issue[15,22-24]. By directly comparing atomic resolution STM images to simulated STM images of defect models, the dominate defects in samples can be unambiguously identified. As shown in previous studies of $Sb_2Te_3$ thin films[23], these true atomic resolution STM images of native defects are crucial to identify them for synthesis optimization of high quality samples.

Among known 3D TIs, $Bi_2Se_3$ is one of the most promising systems because of large band gap (~0.35 eV) and a simple surface state dispersion with Dirac point inside the band gap[25,26]. However, nominal $Bi_2Se_3$ single crystals are *n*-type doped semiconductors because of formation of native defects during synthesis. Earlier studies suggested Se vacancies ($V_{Se}$) were the dominant donors in $Bi_2Se_3$[27,28]. Despite of intensive STM studies, there is no clear identification of $V_{Se}$[15,22,29-31]. Without clear identification and control of native defects, it is difficult to improve the synthesis of $Bi_2Se_3$ single crystals. Some of previous STM works speculated the commonly observed "triangular" shape (protrusion) defects as $V_{Se}$ without clear evidence[22,31,32]. In this letter, we report high resolution STM studies of native defects in various $Bi_2Se_3$ single crystals. Comparison between experiments and first principle simulations of various defects allow us to unambiguously identify native defects and their relative concentrations, including $V_{Se}$, $Bi_{Se}$ antisites and Se interstitial defects ($Se_i$). In particular, we establish that the dominate $V_{Se}$ defect lies in the middle of the Se layer, instead of at the energetically favorable positions at the internal van der Waals (vdW) surfaces. Combined with the presence of Se intercalations, these findings demonstrate the importance of the kinetics over the thermodynamics of defect formation in $Bi_2Se_3$, and possibly in other pnictide chalgegonide TIs. Additionally, our scanning tunneling spectroscopy (STS) mapping demonstrates that as donors $V_{Se}$ defects are also responsible for significant local potential fluctuation[15]. Our systematic STM/STS studies demonstrate that the Fermi level ($E_F$) is lowered with reduction of $V_{Se}$ density via fine control of the stoichiometry. However, too much Se introduces $Se_i$, again resulting in *n*-type doping. With delicate control of stoichiometry and synthesis parameters, we obtained high quality $Bi_2Se_3$ single crystal with $E_F$ ~60 meV below the conduction band minimum, approaching the intrinsic semiconductor limit without compensation doping.

For Se deficient (rich) $Bi_2Se_3$ single crystals, mixtures of high-purity chemicals with various ratio were heated up to 870 ºC (800 ºC) in sealed vacuum quartz tubes for 16 hours, then slowly cooled to 690 ºC, followed by furnace cooling. Stoichiometric $Bi_2Se_3$ single crystals were grown using a slow-cooling method[39]. STM/STS measurements were carried out at 4.5 K in an Omicron LT-STM with base pressure of $1\times10^{-11}$ mbar. Electrochemically etched tungsten tips were characterized on clean Au (111) surface before STM experiments. Single crystals of $Bi_2Se_3$ were inserted to cold STM head immediately after RT cleavage in UHV. STS measurements were performed with standard lock-in technique ($f$ = 455 Hz) with modulation 5~10 mV. Simulated STM images are constructed using the theory of Tersoff and Hamann[33]. The charge densities associated with the defects are obtained self-consistently using density functional theory within the approximation of Perdew-Burke-Ernzerhof[34]. Interactions between the ion cores and valence electrons are described by the projector augmented wave method[35] as implemented in the VASP package[36,37].

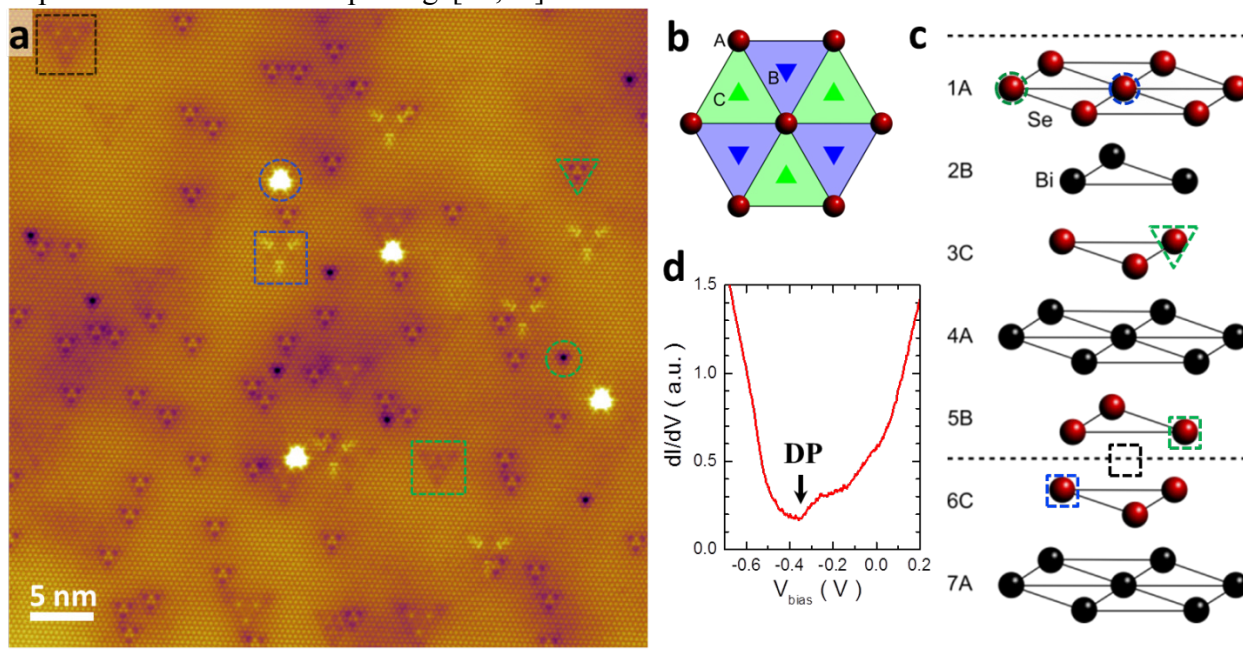

**FIG 1.** (color online) (a) atomically resolved STM image of (001) surface of nominal $Bi_2Se_3$. Tunneling parameter: $U = -0.7$ V, $I = 0.2$ nA. Green circle/triangle/square marks a $V_{Se}$ at 1A/3C/5B site. Blue circle/square marks a $Bi_{Se}$ antisite at 1A/6C sites. Black square marks a triangular defect (Se interstitial defect) at A site. See Fig. 2 for details. (b) cartoon of triangular lattice and definition of A, B and C sites. (c) definition of atomic sites (1A, 2B, 3C, …) in the crystal structure of $Bi_2Se_3$. (d) STS spectrum of nominal $Bi_2Se_3$. The Dirac point energy ($E_D$) is ~ -340 meV.

$Bi_2Se_3$ forms layered structure with vdW bonding between quintuple layers (QL), so it is ideal for surface sensitive probes such as STM. We found atomic corrugation is pronounced at negative bias. Figure 1(a) shows atomically resolved STM image of (001) cleaved surface of a nominal $Bi_2Se_3$ single crystal with slight Se deficiency. STM images with sub-unit cell resolution allows clear identification of defect positions as shown in Fig. 1(b) and 1(c)[23]. There are two kinds of Se site defects, depressions (green) and protrusions (blue), labelled in Fig. 1(c). This is in good agreements with previous first principle theory calculations that low formation energy defects are $V_{Se}$ and $Bi_{Se}$ for Bi rich growth[27,28]. Thus, we only need to determine whether the observed defects are either $V_{Se}$ or $Bi_{Se}$. Prior STS measurements of $Bi_{Se}$ reveal a pronounced local density of states (LDOS) peak near VBM, which is corroborated by recent first principle calculations[22,24,28]. Our STS data of protrusion defects all show pronounced characteristic LDOS peak associated with $Bi_{Se}$, while no LDOS anomaly was found in suppression defects[39]. Therefore, we can assign protrusion defects as $Bi_{Se}$ and suppression defects as $V_{Se}$ as summarized in Figure 2. Our high resolution STM data provide a clear identification of all possible selenium vacancies ($V_{Se}$) in $Bi_2Se_3$, which is further supported by first principle simulation as shown in Fig. 2.

The most popular defects are $V_{Se}$ (green triangle) located at the 3C sites. The estimated density is ~$3\times10^{19}$ cm$^{-3}$, consistent with our STS data (Fig. **1d**) that Fermi level ($E_F$) is ~140 meV above conduction band minimum (CBM) . Another prominent $V_{Se}$ (green circle in Fig. **1a**) is a single depression at 1A site[38]. More subtle $V_{Se}$ (green square) were identified at the 5B site, which is crystallographically equivalent to the 1A site. Consistently, their estimated densities are statistically the same[39]. Note that $V_{Se}$(3C) has not been identified in previous STM studies of $Bi_2Se_3$, probably because of relatively weak atomic corrugation. The observation of dominant $V_{Se}$(3C) are also unexpected because of its high formation energy[27,28]. We speculate that $V_{Se}$(3C) defects formed at high temperature are kinetically trapped during the cooling process of crystal growth, i.e., there is a large energy barrier that prevents Se atoms from diffusing into middle Se layer to annihilate $V_{Se}$(3C) defects.

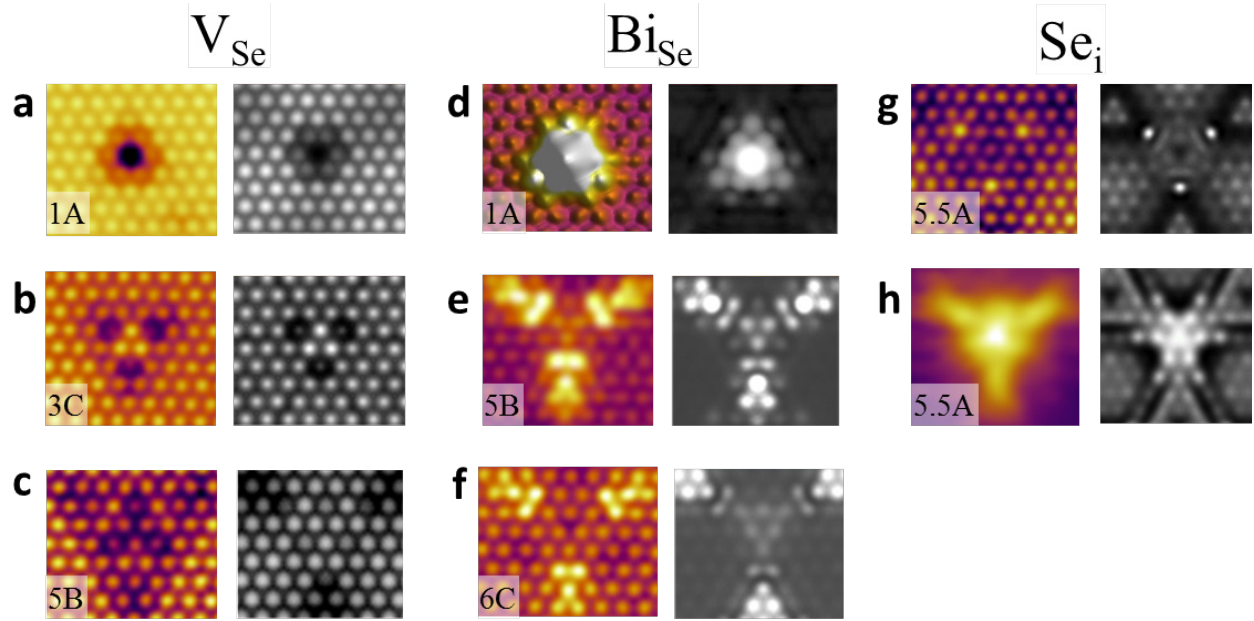

FIG 2. (color online) Comparison between experiment observation of first-principle calculation of native defects in $Bi_2Se_3$. (a), (b), and (c) STM images (left) of Selenium vacancies at 1A, 3C and 5B sites measured with -0.7 V, and the corresponding simulations (right). (d), (e), (f), STM images (left) of $Bi_{Se}$ antisites at 1A, 5B and 6C sites measured with -0.6 V and the corresponding simulations. (g) and (h), Left: STM images of an intercalated Se defect at -0.7 V and +0.5 V, respectively; Right: corresponding simulated STM images.

To substantiate the experimental identification of native defects, we performed simulations of STM images of various $V_{Se}$ and $Bi_{Se}$ defects using first principles. Figure 2 shows the comparisons between experimental results and simulations. The qualitative agreements of defect pattern provide compelling support for our identifications of native defects in $Bi_2Se_3$. In addition to $Bi_{Se}$ on 5B and 6C sites that have been identified by earlier STM works, we identify $Bi_{Se}$ on 1A site, which has not been reported, probably because its pronounced apparent height (~Å) makes it easily confused with surface adsorbed atoms/molecules[39].

Previous STM works associated triangular defects observed at positive bias with Se vacancies[22,32,40]. We also observed triangular defects at positive bias on, as shown in Fig. 2h. Our atomically resolved STM image indicates that it is a defect at the A site ~5-6 atomic layers beneath surface. Clearly its density is much less than that of $V_{Se}$, and thus cannot account for the *n*-type doping in nominal Se deficient $Bi_2Se_3$. The density of triangular defects increases significantly in $Bi_2Se_3$ synthesized in Se rich environment, indicating it is a Se interstitial defect, likely sitting in the vdW gap between QLs (i.e. 5.5A site). This assignment is further substantiated by the positive correlation between the density of triangular defects and that of surface Se atoms, which can be removed by STM tip[39]. Although unambiguous delineation of triangular defects requires further investigation, our STM results unambiguously exclude $V_{Se}$ as a possible candidate.

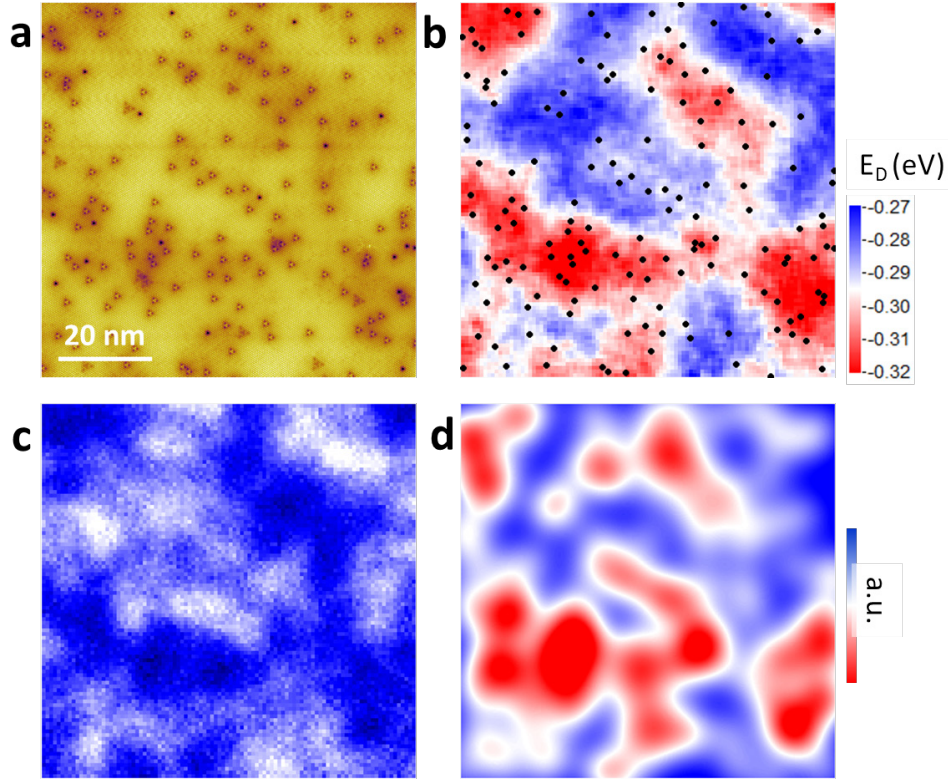

FIG 3 (color online) (a) STM topography image of $Bi_2Se_3$ showing randomly distributed $V_{Se}$ defects. (b) Dirac point energy ($E_D$) map of the same location as a. Black dots are the positions of $V_{Se}$ defects in (a). There is a clear positive correlation between local $V_{Se}$ defect density and local $E_D$. (c) local conduction ($dI/dV$) map at -500 meV showing the local potential fluctuation. (d) Simulated defect influence map due to $V_{Se}$ [37].

As donors, $V_{Se}$ are charged defects that influence the local potential. Previous STM studies observed pronounced fluctuation of local potential and $E_D$ of surface states in both doped and undoped $Bi_2Se_3$ single crystals[15,22]. Yet, the microscopic origin of potential fluctuation has not been clarified. Our STM results demonstrate that $V_{Se}$ is responsible for the spatial fluctuation of potential (and $E_D$). Figure 3(a) shows a STM topographic image of randomly distributed $V_{Se}$ defects in a nominally Se-deficient $Bi_2Se_3$ single crystal. STS measurements at each pixel allow us to extract LDOS and $E_D$ maps. Fig. 3(b) shows spatial fluctuation $E_D$ at each location in this area[39]. Consistently, the same potential fluctuation pattern can be found in Fig. 3(c), $dI/dV$ map at -500 meV. Clearly there is a positive correlation between local $E_D$ and the local density of $V_{Se}$ defects, illustrated by the simulated defect influential map in Fig. 3(d)[39]. Therefore, the presence of significant amount of $V_{Se}$ defects not only causes substantial bulk conduction via n-type doping, but also induces a significant spatial fluctuation of electric potential that is detrimental for the topological transport properties.

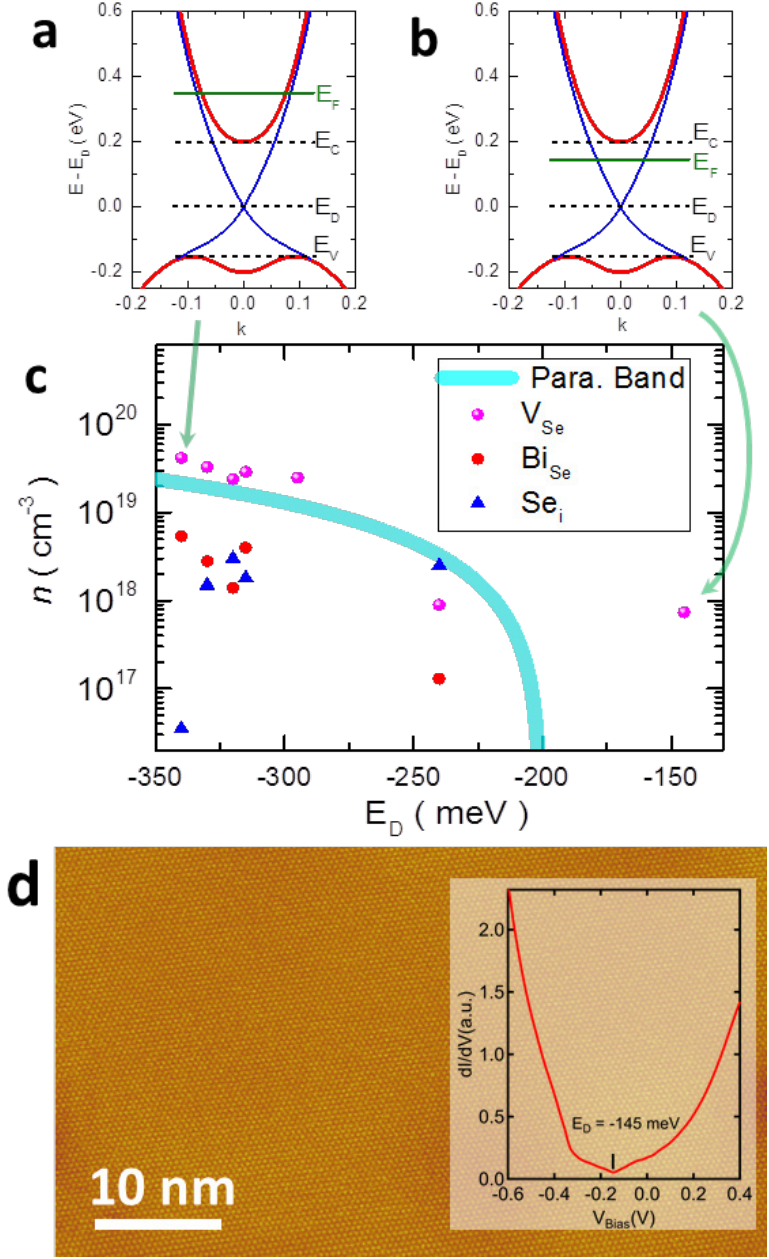

FIG 4. (color online) (a)/(b), cartoons of $E_F$ relative to band structure in typical/best $Bi_2Se_3$ samples. (c) a summary diagram ($n$ vs. $E_D$) of various $Bi_2Se_3$ crystals with various synthesis conditions. Assuming $V_{Se}$ and $Se_i$ are donors, the cyan line shows the electron density estimated from parabolic band ($m^* \approx 0.15\ m_e$)[41]. (d) Typical STM topography of sample with extremely low defect density. Inset: corresponding STS showing that $E_F$ is below CBM.

With clear identification of native defects in $Bi_2Se_3$, one can correlate the defect population with growth conditions to gain insights on synthesis. Figure 4(c) shows defect densities ($n$) and Dirac point energy ($E_D$) measured on various $Bi_2Se_3$ single crystals. Clearly, $E_D$ increases systematically with decreasing $n$ of $V_{Se}$, in good agreement with estimated carriers density from simple 3D parabolic band (cyan line)[41,42]. The systematic behavior of $n$ vs. $E_D$ further corroborates our defect identifications. In addition, the reduction of $V_{Se}$ (and $Bi_{Se}$) density is correlated with increase of molar fraction of Se in the starting materials, in agreement with previous studies and formation energy calculations[27,41,42]. However, our results indicate that Se-rich condition results in a substantial increase of density of $Se_i$ (triangular defects) instead of $Se_{Bi}$ antisites predicted by defect thermodynamics[27]. Our STM data suggest $Se_i$ are also electron donors, which explains the fact that Se-rich synthesis still results in *n*-type doping, as shown in the sample with $E_D \approx$ -240 meV.

With the clear understanding of native defects in $Bi_2Se_3$, we are in the position to fine tune the synthesis recipe to improve sample quality. With delicate control of stoichiometry of staring materials (Bi:Se = 2:3), the tendency of forming $V_{Se}$ and $Se_i$ are greatly suppressed. Crystals were slowly cooled to room temperature during growth to avoid kinetically trapped $V_{Se}$ in the middle Se layer. Figure 4(d) shows a typical STM image of our best sample, where the defect density is very low ($<10^{18}$ cm$^{-3}$) with negligible potential fluctuation. Furthermore, our STS measurements (Fig. 4e) suggest the Dirac point energy is -145 meV, indicating the Fermi level is ~60 meV below CBM[41]. Although the system is still slightly *n*-doped, further optimization of synthesis condition would likely result in samples closer to the intrinsic limit.

In summary, we have unambiguously identified the native defects, especially $V_{Se}$ in $Bi_2Se_3$ single crystals by combining systematic high resolution STM/STS measurements and first-principle simulations. Our results reveal that the native defect concentrations in $Bi_2Se_3$ do not reflect their thermodynamic formation energies, but instead are highly influenced by the kinetics of growth. This is seen not only in the dominate $V_{Se}$ residing in the middle Se layer, but also in the presence of significant Se interstitial defects (despite their high formation energy) and the absence of the low energy substitutional Se defect ($Se_{Bi}$). In addition, our results suggest that the "triangular defects" seen in previous works are likely Se interstitial defects ($Se_i$) which are dominant in Se-rich growth. These findings emphasize the importance of both strict stoichiometry and kinetic control of defect formation in synthesis, leading to improved synthesis of high quality single crystals with low (charge) defect density with minimal potential fluctuation. With Fermi level below the CBM, our results suggest a viable route toward intrinsic limit of topological insulators for potential applications, which may shed new light for synthesis of other technological relevant functional materials[1].

**Acknowledgement**

We are grateful to D. Vanderbilt for helpful discussions. JD and WW thanks W. Zhang for assistance on data analysis. The STM work at Rutgers was supported by NSF Grants No. DMR-0844807 and No. DMR-1506618. The crystal-growth work at Rutgers was supported by the NSF under Grant No. NSF-DMREF-1233349. SBZ was supported by the Department of Energy under Grant No. DE-SC0002623. DW was supported by NSF under grant No. NSF-1542798. The supercomputer time was provided by the CCNI at RPI and NERSC under DOE Contract No. DE-AC02-05CH11231.

# References

[1] C. Freysoldt, B. Grabowski, T. Hickel, J. Neugebauer, G. Kresse, A. Janotti, and C. G. Van de Walle, Reviews of Modern Physics **86**, 253 (2014).
[2] P. M. Voyles, D. A. Muller, J. L. Grazul, P. H. Citrin, and H. J. L. Gossmann, Nature **416**, 826 (2002).
[3] Y. Ando, Journal of the Physical Society of Japan **82**, 102001 (2013).
[4] X.-L. Qi and S.-C. Zhang, Reviews of Modern Physics **83**, 1057 (2011).
[5] M. Z. Hasan and J. E. Moore, Annual Review of Condensed Matter Physics **2**, 55 (2011).
[6] M. Z. Hasan and C. L. Kane, Reviews of Modern Physics **82**, 3045 (2010).
[7] L. Fu, C. L. Kane, and E. J. Mele, Phys. Rev. Lett. **98**, 106803 (2007).
[8] S. Oh, Science **340**, 153 (2013).
[9] C.-Z. Chang *et al.*, Science **340**, 167 (2013).
[10] C. Z. Chang *et al.*, Nat Mater **14**, 473 (2015).
[11] J. S. Zhang *et al.*, Nature Communications **2**, 574 (2011).
[12] Z. Ren, A. A. Taskin, S. Sasaki, K. Segawa, and Y. Ando, Phys. Rev. B **82**, 241306(R) (2010).
[13] D. Hsieh *et al.*, Nature **460**, 1101 (2009).
[14] D. Kim, S. Cho, N. P. Butch, P. Syers, K. Kirshenbaum, S. Adam, J. Paglione, and M. S. Fuhrer, Nat Phys **8**, 459 (2012).
[15] H. Beidenkopf, P. Roushan, J. Seo, L. Gorman, I. Drozdov, Y. S. Hor, R. J. Cava, and A. Yazdani, Nature Phys. **7**, 939 (2011).
[16] Y. L. Chen *et al.*, Science **325**, 178 (2009).
[17] K. Hoefer, C. Becker, D. Rata, J. Swanson, P. Thalmeier, and L. H. Tjeng, Proc Natl Acad Sci U S A **111**, 14979 (2014).
[18] K. Hoefer, C. Becker, S. Wirth, and L. Hao Tjeng, AIP Advances **5**, 097139 (2015).
[19] T. Bathon, S. Achilli, P. Sessi, V. A. Golyashov, K. A. Kokh, O. E. Tereshchenko, and M. Bode, Adv Mater **28**, 2183 (2016).
[20] C. G. Van de Walle and J. Neugebauer, Journal of Applied Physics **95**, 3851 (2004).
[21] D. A. Drabold and S. K. Estreicher, *Theory of Defects in Semiconductors* (Springer-Verlag, Berlin Heidelberg, 2007), Vol. 104, Topics in Applied Physics.
[22] C. Mann, D. West, I. Miotkowski, Y. P. Chen, S. Zhang, and C.-K. Shih, Nat Commun **4**, 2277 (2013).
[23] Y. Jiang *et al.*, Phys. Rev. Lett. **108**, 066809 (2012).
[24] S. Urazhdin, D. Bilc, S. H. Tessmer, S. D. Mahanti, T. Kyratsi, and M. G. Kanatzidis, Phys. Rev. B **66**, 161306 (2002).
[25] H. J. Zhang, C. X. Liu, X. L. Qi, X. Dai, Z. Fang, and S. C. Zhang, Nature Phys. **5**, 438 (2009).
[26] Y. Xia *et al.*, Nature Phys. **5**, 398 (2009).
[27] D. West, Y. Y. Sun, H. Wang, J. Bang, and S. B. Zhang, Phys. Rev. B **86**, 121201 (2012).
[28] L.-L. Wang, M. Huang, S. Thimmaiah, A. Alam, S. L. Bud'ko, A. Kaminski, T. A. Lograsso, P. Canfield, and D. D. Johnson, Phys. Rev. B **87**, 125303 (2013).
[29] S. Urazhdin, D. Bilc, S. D. Mahanti, S. H. Tessmer, T. Kyratsi, and M. G. Kanatzidis, Phys. Rev. B **69**, 085313 (2004).
[30] S. Jia, H. Beidenkopf, I. Drozdov, M. K. Fuccillo, J. Seo, J. Xiong, N. P. Ong, A. Yazdani, and R. J. Cava, Phys. Rev. B **86**, 165119 (2012).
[31] Y. S. Hor, A. Richardella, P. Roushan, Y. Xia, J. G. Checkelsky, A. Yazdani, M. Z. Hasan, N. P. Ong, and R. J. Cava, Phys. Rev. B **79**, 195208 (2009).
[32] Z. Alpichshev, R. R. Biswas, A. V. Balatsky, J. G. Analytis, J. H. Chu, I. R. Fisher, and A. Kapitulnik, Phys. Rev. Lett. **108**, 206402 (2012).
[33] J. Tersoff and D. R. Hamann, Phys. Rev. B **31**, 805 (1985).
[34] J. P. Perdew, K. Burke, and M. Ernzerhof, Phys. Rev. Lett. **77**, 3865 (1996).

[35]	P. E. Blochl, Phys. Rev. B **50**, 17953 (1994).
[36]	G. Kresse and J. Furthmuller, Phys. Rev. B **54**, 11169 (1996).
[37]	G. Kresse and D. Joubert, Phys. Rev. B **59**, 1758 (1999).
[38]	P. Cheng *et al.*, Phys. Rev. Lett. **105**, 076801 (2010).
[39]	See supplemental materials.
[40]	Y. S. Hor *et al.*, Phys. Rev. Lett. **104**, 057001 (2010).
[41]	J. G. Analytis, J. H. Chu, Y. L. Chen, F. Corredor, R. D. McDonald, Z. X. Shen, and I. R. Fisher, Phys. Rev. B **81**, 205407 (2010).
[42]	N. P. Butch, K. Kirshenbaum, P. Syers, A. B. Sushkov, G. S. Jenkins, H. D. Drew, and J. Paglione, Phys. Rev. B **81**, 241301(R) (2010).